\documentstyle[psfig]{mn}

\footnotesize
\newdimen\digitwidth    
\setbox0=\hbox{\rm0}
\digitwidth=\wd0
\catcode`!=\active
\def!{\kern\digitwidth}
\normalsize

\title[Characteristic signature of preferred frame effects in binary pulsars]{ 
A characteristic observable signature of preferred frame effects in 
relativistic binary pulsars }

\author[N.~Wex \& M.~Kramer]
{
N.~Wex,
M. Kramer\thanks{Email: Michael.Kramer@manchester.ac.uk}
\\
University of Manchester,
Jodrell Bank Observatory, Macclesfield, Cheshire, SK11~9DL, UK
}

\date{}
\begin{document}

\maketitle
\newcommand{\setthebls}{
}

\setthebls


\begin{abstract} 
  In this paper we develop a consistent, phenomenological methodology to
  measure preferred-frame effects (PFEs) in binary pulsars that exhibit a high
  rate of periastron advance. We show that in these systems the existence of a
  preferred frame for gravity leads to an observable characteristic
  `signature' in the timing data, which uniquely identifies this effect. We
  expand the standard Damour-Deruelle timing formula to incorporate this
  `signature' and show how this new PFE timing model can be used to either
  measure or constrain the parameters related to a violation of the local
  Lorentz invariance of gravity in the strong internal fields of neutron
  stars. In particular, we demonstrate that in the presence of PFEs we expect
  a set of the new timing parameters to have a unique relationship that can be
  measured and tested incontrovertibly. This new methodology is applied to the
  Double Pulsar, which turns out to be the ideal test system for this kind of
  experiments.  The currently available dataset allows us only to study the
  impact of PFEs on the orbital precession rate, $\dot\omega$, providing
  limits that are, at the moment, clearly less stringent than existing limits
  on PFE strong-field parameters.  However, simulations show that the
  constraints improve fast in the coming years, allowing us to study all new
  PFE timing parameters and to check for the unique relationship between them.
  Finally, we show how a combination of several suitable systems in a {\em PFE
  antenna array}, expected to be availabe for instance with the
  Square-Kilometre-Array (SKA), provides full sensitivity to possible
  violations of local Lorentz invariance in strong gravitational fields in all
  directions of the sky. This PFE antenna array may eventually allow us to
  determine the direction of a preferred frame should it exist.
\end{abstract}


\begin{keywords}
gravitation, pulsars:general, pulsars:individual:PSR J0737$-$3039
\end{keywords}


\section{Introduction}

The theory of general relativity (GR) has so far passed all experimental tests
with flying colours (Stairs~2003\nocite{sta03}, Will~2006\nocite{wil06},
Kramer et al.~2006\nocite{ksm+06}). Nevertheless, GR may not be the final word
in our understanding of gravity and indeed, alternative theories of gravity
exist, predicting deviations from GR in various possible ways.

Some theories of gravity predict that the Universe's global matter
distribution selects a preferred rest frame for local gravitational physics.
In such theories the outcome of gravitational experiments depends on the motion
of the laboratory with respect to this preferred frame. In particular,
theories in which gravity is partially mediated by a vector field or a second
tensor field are known to exhibit such preferred-frame effects whose strength
is determined by cosmological matching parameters.

In the post-Newtonian limit, such preferred-frame effects are described by two
phenomenological parameters, the parameterised post-Newtonian (PPN) parameters
$\alpha_1$ and $\alpha_2$. In GR these two parameters are zero.  Solar system
experiments have already tightly constrained these two PPN parameters. For
details see Will~(1993\nocite{wil93}, 2006\nocite{wil06}) and references
therein. On the other hand, limits obtained in the weak gravitational fields
of the solar system cannot rule out effects that only become significant in
strong gravitational fields. For the well-motivated scalar-tensor theories,
this has been demonstrated in a series of papers by Damour and
Esposito-Far{\`e}se (1992a\nocite{de92a}, 1993\nocite{de93},
1996a\nocite{de96a}, 1996b\nocite{de96b}).

In order to describe preferred-frame effects in binary pulsars, Damour and
Esposito-Far{\`e}se (1992b)\nocite{de92b} introduced a non-boost-invariant
Lagrangian that includes two strong field parameters $\hat\alpha_1$ and
$\hat\alpha_2$, capable of accounting for strong-field violations of a
local Lorentz invariance of gravity. In the weak-field limit these two
strong-field parameter are equal to $\alpha_1$ and $\alpha_2$, respectively.
Concentrating on the contribution by $\hat\alpha_1$, Damour and
Esposito-Far{\`e}se show that $\hat\alpha_1$ has the effect of inducing
secular variations in the orbital eccentricity and in the orientation of the
pulsar orbit, i.e. the location of periastron and the orientation of the
orbital plane. These variations depend on the magnitude and direction of the
velocity with respect to the preferred frame. Their calculations for
small-eccentricity orbits shows that the eccentricity oscillates in a specific
way between a maximum and a minimum value, with the period of the relativistic
precession of periastron. From this, using probabilistic considerations, they
were able to derive an upper limit of $|\hat\alpha_1| < 5 \times 10^{-4}$ with
a 90\% confidence level. After the discovery of a number of new
small-eccentricity binaries, Wex~(2000)\nocite{wex00} extended this method in
statistically combining multiple systems, by this taking care of a potential
selection effect when simply picking the system with the most favourable
parameter combination. His analysis yielded the slightly improved limit of
$|\hat\alpha_1| < 1.2 \times 10^{-4}$ with a 95\% confidence level.

In a recent paper Bailey and Kosteleck{\'y}~(2006)\nocite{bk06} study the
violation of local Lorentz invariance within an effective field theory called
the standard model extension (SME). Besides a number of solar system tests,
they use their formalism to investigate the measurability of SME
preferred-frame effects in binary pulsars. However, as already pointed out by
Bailey and Kosteleck{\'y}, their formalism does not account for strong field
effects. In addition, as we will demonstrate in the course of this paper,
their usage of the PSR~B1913+16 timing results does not provide a consistent
test.

In this paper we present a new method for testing preferred-frame effects 
related to strong gravitational fields, which uses binary pulsars with a high 
rate of periastron advance. Our method provides independent and simultaneous 
tests for both $\hat \alpha_1$ and $\hat\alpha_2$.  While the tests of Damour 
and Esposito-Far{\`e}se~(1992b)\nocite{de92b} and Wex~(2000)\nocite{wex00}, by 
their probabilistic considerations, can only provide upper limits for 
preferred-frame effects, the method introduced in this paper is even capable of 
detecting a preferred frame in the universe, if it exists.

We apply our new method to the PSR~J0737$-$3039 binary pulsar system
(Burgay~et~al.~2003\nocite{bdp+03}; Lyne~et~al.~2004\nocite{lbk+04}), which is
not only the only Double Pulsar system known, but which also exhibits the
highest rate of periastron advance, $\dot{\omega} = d\omega/dt$, of any binary
pulsar known. The precisely measured value of $\dot{\omega}=16.89947(68)$ deg
yr$^{-1}$ (Kramer~et~al.~2006\nocite{ksm+06}) is more than four times larger
than the value measured for the Hulse-Taylor pulsar PSR~B1913+16 (Weisberg \&
Taylor 2002\nocite{wt02}). While the Double Pulsar has been used to provide
the so far most stringent test of GR in the strong field regime, we show here
that the large periastron advance makes it also a unique testbed to actually
measure a possible violation of the Lorentz invariance for strong gravitational
fields, which may occur in alternative theories of gravity.

The plan of the paper is as follows. In Section~\ref{motion}, based on a
theory-independent framework of Will~(1993), we study in full the motion of a
compact binary system in theories of gravity that are non-boost-invariant. We
calculate the secular evolution, caused by preferred-frame effects, for all
binary parameters, in order to show that the combination of these periodic
parameter changes exhibits a unique `signature' of a preferred frame. In
Section~\ref{timing} we investigate the impact of such time-dependencies on
the high-precision timing measurements of binary pulsars, and how this can be
used to either measure or constrain strong-field preferred-frame effects. In
Section~\ref{psr0737} we apply our method to the Double Pulsar to obtain
limits for preferred-frame effects from this system. Finally, we present
results of simulations that show how the precision of this test will improve
within the next couple of years, and we investigate the possibility of
combining observations of several binary pulsar systems, to obtain a full sky
coverage with high sensitivity.


\section{Preferred-frame effects in the motion of compact binaries}
\label{motion}

In this first part we study the motion of compact binary systems in theories
of gravity that have a preferred frame of reference. We derive the equations
of motion from a general formalism introduced by Will~(1993) which includes
strong field contributions using generic strong-field parameters.  As we do
not impose boost-invariance, the resulting equations of motion contain terms
that depend on the motion of the binary system with respect to a preferred
frame. Hence, our computations generalize the earlier work by Nordtvedt and
Will~(1972)\nocite{nw72} and Damour and Esposito-Far{\`e}se~(1992b) to
describe binary systems and their orbital parameters with arbitrary
eccentricities. Using the method of perturbations of osculating orbital
elements it is shown that the rate of relativistic precession of periastron,
the orbital eccentricity, and the orbital period should change in a
characteristic way over time if a binary system moves relative to a preferred
frame of reference.

\subsection{Binary dynamics in the generalised EIH formalism}

When dropping constant terms and rescaling the masses, the semi-conservative,
generalised Einstein-Infeld-Hoffmann (EIH) Lagrangian of Will~(1993) reads for
a system consisting of two compact objects
\begin{equation}
  L = L^{(0)} + L^{(1)}/c^2
\end{equation}
where
\begin{equation}
  L^{(0)} =   \mbox{$\frac{1}{2}$} m_p v_p^2 
            + \mbox{$\frac{1}{2}$} m_c v_c^2 
            + \frac{{\cal G} G m_p m_c}{r} 
\end{equation}
and
\begin{eqnarray}
&&\hspace{-5.0ex} L^{(1)} = 
    \mbox{$\frac{1}{8}$} {\cal A}_p m_p v_p^4 
  + \mbox{$\frac{1}{8}$} {\cal A}_c m_c v_c^4 
  \nonumber\\
&&\hspace{-3.0ex}
  + \frac{G m_p m_c}{2r}\left[ 3{\cal B}(v_c^2 + v_p^2) 
  - 7{\cal C}({\bf v}_p \cdot {\bf v}_c)
  -  {\cal E}({\bf v}_p \cdot \hat{\bf n})({\bf v}_c \cdot \hat{\bf n})\right] 
  \nonumber\\
&&\hspace{-3.0ex}
  - \frac{G m_p m_c}{2r^2}\left[ m_p{\cal D}_c + m_c {\cal D}_p \right] \:. 
\end{eqnarray}
$m_p$ and $m_c$ are the inertial masses of pulsar and companion, respectively,
$r=|{\bf r}|=|{\bf x}_p-{\bf x}_c|$ is the coordinate distance between pulsar 
and companion, $\hat{\bf n}\equiv{\bf r}/r$, ${\bf v}_p$ and ${\bf v}_c$ are 
the coordinate velocities of pulsar and companion, respectively, and $G$ is the 
Newtonian constant of gravity.

The coefficients ${\cal A}_p$, ${\cal A}_c$, ${\cal B}$, ${\cal C}$, 
${\cal D}_p$, ${\cal D}_c$, ${\cal E}$, and ${\cal G}$ are functions of the
parameters of the chosen theory of gravity and of the structure of each body.
They account for contributions of the highly relativistic interior of the
neutron stars to the binary dynamics.  In particular, it is assumed that these
coefficients are independent of the interbody distance in the binary system.
In this paper we will denote them as `strong gravity coefficients'. The strong
gravity coefficients ${\cal B}$, ${\cal C}$, ${\cal E}$, and ${\cal G}$ are
symmetric under interchange of $p \leftrightarrow c$. For instance, in the
fully conservative version of Rosen's bimetric theory ${\cal G} = 1 -
\frac{4}{3}s_p s_c$, where $s_p$ and $s_c$ are the `first sensitivities'
($\equiv -\partial \ln m / \partial \ln G \sim [\mbox{gravitational binding
  energy}]/[\mbox{mass}]$) of pulsar and companion, respectively. In GR, these
parameters are unity and, therefore, the compactness of a body does not have
an impact on its orbital motion, a property of GR known as the `effacement' of
internal structure (see the discussion in Damour~1987\nocite{dam87}).

We note that combining the PFE-Terms of the Lagrangian of Damour and
Esposito-Far{\`e}se~(1992b) and the generalised conservative Lagrangian of
Damour and Esposito-Far{\`e}se~(1992a) yields a Lagrangian that is equivalent
to the generalised Lagrangian of Will~(1993) if one sets the ${\cal A}$-terms
in Will's Lagrangian equal to in unity. As it turns out that these 
${\cal A}$-terms play an important role in the violation of Lorentz invariance, 
we use Will's Lagrangian for our calculations.

In contrast to Will~(1993), we do not impose post-Galilean invariance when
calculating the equations of motion, in order to include gravitational
theories that have a preferred frame of reference and, therefore, are not 
post-Galilean invariant. Hence, the equations of motion for pulsar and 
companion in the preferred frame are derived from the Euler-Lagrange equations
\begin{equation} \label{eq:EulerLagrange}
  \frac{d}{dt}\frac{\partial L}{\partial {\bf v}_p}
            - \frac{\partial L}{\partial {\bf x}_p} = 0
\quad \mbox{and} \quad
  \frac{d}{dt}\frac{\partial L}{\partial {\bf v}_c}
            - \frac{\partial L}{\partial {\bf x}_c} = 0 \:,
\end{equation}
without any further restrictions on the `strong gravity coefficients' of $L$.
The relative acceleration ${\bf a} = {\bf a}_p - {\bf a}_c$ for a binary pulsar 
system at rest with respect to the preferred frame is then found to be
\begin{equation}
  {\bf a} = {\bf a}^{(0)} + {\bf a}^{(1)}/c^2 \:,
\end{equation}
where
\begin{equation}
  {\bf a}^{(0)} = -\frac{m^\ast\hat{\bf n}}{r^2} \:,
\end{equation}
\begin{eqnarray}
&&\hspace{-5.0ex} {\bf a}^{(1)} = 
  \frac{m^\ast\hat{\bf n}}{r^2} \left\{\left[1 + 3{\cal B}^\ast 
       + (2+\alpha_1^\ast)\xi_p\xi_c \frac{}{}\right.
  \right. \nonumber\\
&&\hspace{13.0ex} \left.\left. 
  -(1-{\cal D}_c^\ast)\xi_p - (1-{\cal D}_p^\ast)\xi_c
  \frac{}{}\right] \frac{m^\ast}{r} 
  \right. \nonumber\\
&&\hspace{9.0ex} \left.
  - \left[
    \mbox{$\frac{1}{2}$}(3{\cal B}^\ast - 1)
  + \mbox{$\frac{1}{2}$}(6 + \alpha_1^\ast + \alpha_2^\ast)\xi_p\xi_c \frac{}{}
  \right.\right. \nonumber\\
&&\hspace{13.0ex} \left.\left.
  + \mbox{$\frac{1}{2}$}(1 - {\cal A}_c) \xi_p^3
  + \mbox{$\frac{1}{2}$}(1 - {\cal A}_p) \xi_c^3 \frac{}{}\right] v^2 
  \right. \nonumber\\
&&\hspace{9.0ex} \left.
  + \mbox{$\frac{3}{2}$}(1+\alpha_2^\ast)\xi_p\xi_c ({\bf v}\cdot\hat{\bf n})^2 
    \frac{}{}\right\} 
\nonumber\\
&&\hspace{1.0ex} 
  + \frac{m^\ast{\bf v}}{r^2} \left[
    1 + 3{\cal B}^\ast - (2 - \alpha_1^\ast + \alpha_2^\ast) \xi_p\xi_c 
\frac{}{} \right.\nonumber\\ 
&&\hspace{13.0ex}\left.
  - (1 - {\cal A}_c) \xi_p^3 - (1 - {\cal A}_p) \xi_c^3 \frac{}{}\right]
  ({\bf v}\cdot\hat{\bf n}) \:,
\end{eqnarray}
where $m^\ast={\cal G}G(m_p+m_c)$, $\xi_p=m_p/m$, $\xi_c=m_c/m=1-\xi_p$,
${\cal B}^\ast= {\cal B}/{\cal G}$, ${\cal C}^\ast={\cal C}/{\cal G}$, ${\cal
D}_p^\ast= {\cal D}_p/{\cal G}^2$, ${\cal D}_c^\ast={\cal D}_c/{\cal G}^2$,
${\cal E}^\ast = {\cal E}/{\cal G}$,
and further
\begin{equation}
  \alpha_1^\ast = {\cal E}^\ast + 7{\cal C}^\ast - 6{\cal B}^\ast - 2 \:,\\
\end{equation}
\begin{equation}
  \alpha_2^\ast = {\cal E}^\ast - 1 \:. 
\end{equation}
Note, in GR $\alpha_1^\ast \equiv \alpha_2^\ast \equiv 0$. Further,
$\alpha_1^\ast$ and $\alpha_2^\ast$ are proportional to the PFE parameters
$\hat\alpha_1$ and $\hat\alpha_2$ of Damour and Esposito-Far{\`e}se~(1992b),
respectively: $\alpha_i^\ast = {\cal G}\hat\alpha_i$.

The `Newtonian' part of the above equations of motion, ${\bf a}^{(0)}$, has the 
well-known Keplerian solution
\begin{equation}
  {\bf r} = r({\bf e}_X\cos\phi+{\bf e}_Y\sin\phi ) \:, \qquad
  r = \frac{p}{1+e\cos\phi}
\end{equation}
\begin{equation}
  r^2\dot\phi = (m^\ast p)^{1/2} \:, 
\end{equation}
where
\begin{equation}
  p = a(1-e^2) \quad\mbox{and}\quad (P_b/2\pi)^2 = a^3/m^\ast \:.
\end{equation}
$P_b$ is the orbital period of the binary system and $e$ the eccentricity of
its orbit.

The post-Newtonian terms, ${\bf a}^{(1)}/c^2$, produce a secular advance of
periastron given by
\begin{equation}\label{eq:omegadot1}
  \langle\dot\omega\rangle = \frac{6\pi m^\ast}{c^2 p P_b}\hat{\cal P} \:,
\end{equation}
where
\begin{eqnarray}
  \hat{\cal P} = {\cal B}^\ast 
  + \mbox{$\frac{1}{6}$}(1 - {\cal D}_p^\ast) \xi_c 
  + \mbox{$\frac{1}{6}$}(1 - {\cal D}_c^\ast) \xi_p 
\hspace{7.5em} && \nonumber\\
  + \mbox{$\frac{1}{6}$}(2\alpha_1^\ast - \alpha_2^\ast)\xi_p\xi_c
  - \mbox{$\frac{1}{6}$} (1 - {\cal A}_p)\xi_c^3 
  - \mbox{$\frac{1}{6}$} (1 - {\cal A}_c)\xi_p^3 \:.
&&
\end{eqnarray}

\subsection{Preferred-frame effects in the relativistic binary motion}

In the previous Section we have given results for a binary-pulsar system that
is at rest with respect to the preferred frame of reference. In this Section 
we investigate the dynamics of a binary system that moves relative to the 
preferred frame with velocity ${\bf w}$. As a matter of convenience, we will 
use a frame that is comoving with the binary system. The equations of motion 
in the comoving frame can be derived by imposing a post-Galilean transformation 
(Chandrasekhar and Contopoulos 1967\nocite{cc67}) on the equations of motion 
that result from the Euler-Lagrange equations~(\ref{eq:EulerLagrange}) of the 
previous Section. For the relative acceleration, as expressed in the comoving 
frame, one finds
\begin{equation}
  {\bf a} = {\bf a}^{(0)} + {\bf a}^{(1)}/c^2 + {\bf a}^{(w)}/c^2 \:,
\end{equation}
where
\begin{eqnarray}\label{eq:aw}
&&\hspace{-5.0ex} {\bf a}^{(w)} = 
  \nonumber\\
&&\hspace{-3.0ex}   
  - \frac{m^\ast\hat{\bf n}}{2r^2} \left[ 
    \alpha_1^\ast (\xi_p-\xi_c)({\bf w}\cdot{\bf v}) 
  - 3\alpha_2^\ast           ({\bf w}\cdot\hat{\bf n})^2 \right.\nonumber\\
&&\hspace{3.0ex}\left.
  + (1 - {\cal A}_c)\xi_p w^2 
  + (1 - {\cal A}_p)\xi_c w^2 \right.\nonumber\\
&&\hspace{3.0ex}\left.
  - (1 - {\cal A}_c)\xi_p^2 ({\bf w}\cdot{\bf v})
  + (1 - {\cal A}_p)\xi_c^2 ({\bf w}\cdot{\bf v}) \right]
  \nonumber\\
&&\hspace{-3.0ex}  
  + \frac{m^\ast{\bf w}}{2r^2} \left[ 
    \alpha_1^\ast (\xi_p-\xi_c)({\bf v}\cdot\hat{\bf n}) 
  - 2\alpha_2^\ast           ({\bf w}\cdot\hat{\bf n}) \right.\nonumber\\
&&\hspace{3.0ex}\left.
  - 2(1 - {\cal A}_c)\xi_p ({\bf w}\cdot\hat{\bf n})
  - 2(1 - {\cal A}_p)\xi_c ({\bf w}\cdot\hat{\bf n}) \right.\nonumber\\
&&\hspace{3.0ex}\left.
  + 2(1 - {\cal A}_c)\xi_p^2 ({\bf v}\cdot\hat{\bf n})
  - 2(1 - {\cal A}_p)\xi_c^2 ({\bf v}\cdot\hat{\bf n}) \right]
  \nonumber\\
&&\hspace{-3.0ex}
  - \frac{m^\ast{\bf v}}{r^2} \: \left[  
  (1 - {\cal A}_c)\xi_p^2 ({\bf w}\cdot\hat{\bf n}) 
  - (1 - {\cal A}_p)\xi_c^2 ({\bf w}\cdot\hat{\bf n}) \right]\:,
\end{eqnarray}

The variations of the orbital parameters of a binary pulsar caused by the
preferred-frame acceleration ${\bf a}^{(w)}/c^2$ can now be calculated by the
standard technique of perturbations of osculating orbital elements. Using the
notation of Damour and Taylor~(1992)\nocite{dt92}, one finds, when averaging
over one full orbit, for the change in the longitude of periastron
\begin{eqnarray}\label{eq:deltaomega_w}
  \Delta\omega^{(w)} =
     \frac{\pi m^\ast}{c^2 p} \left[ \mbox{$\frac{1}{2}$} \hat{\cal Q}_1
     \left(\frac{1}{e} - eF_e^2\right)
     \left(\frac{w}{v_0}\right)\sin\psi\sin\chi 
  \right. \hspace{2em} && \nonumber\\ \left. 
     - \hat{\cal Q}_2 F_e^2
       \left(\frac{w}{v_0}\right)^2\sin^2\!\psi\cos 2\chi \right]        
     - \Delta\Omega\cos i \:, &&
\end{eqnarray}
for the change in the longitude of the ascending node
\begin{eqnarray}
  \Delta\Omega^{(w)} = \frac{\pi m^\ast}{c^2 p\sin i} F_e 
    \left[ \hat{\cal Q}_1 e 
    \left(\frac{w}{v_0}\right)\cos\psi\cos\omega
  \right. \hspace{4em} && \nonumber\\ \left. 
    -\hat{\cal Q}_2
    \left(\frac{w}{v_0}\right)^2\sin 2\psi
    \left(\frac{\cos\chi\sin\omega }{\sqrt{1-e^2}}+\sin\chi\cos\omega\right)
    \right] \:,
\end{eqnarray}
for the change in the projected semi-major axis
\begin{eqnarray}\label{eq:deltax_w}
  \frac{ \Delta x^{(w)}}{x} = -\frac{\pi m^\ast}{c^2 p\tan i} F_e 
    \left[ \hat{\cal Q}_1 e
    \left(\frac{w}{v_0}\right)\cos\psi\sin\omega \frac{}{}
  \right. \hspace{3em} && \nonumber\\ \left. 
     +\hat{\cal Q}_2 \left(\frac{w}{v_0}\right)^2\sin 2\psi
     \left(\frac{\cos\chi\cos\omega}{\sqrt{1-e^2}} 
     -\sin\chi\sin\omega\right) \right]  \:, &&
\end{eqnarray}
for the change in the eccentricity
\begin{eqnarray}\label{eq:deltae_w}
   \Delta e^{(w)} = \frac{\pi m^\ast}{c^2 p}(1+e^2)F_e
     \left[  \hat{\cal Q}_1 \left(\frac{w}{v_0}\right)\sin\psi\cos\chi
   \right. \hspace{3em} && \nonumber\\ \left.
            + \hat{\cal Q}_2 \frac{eF_e}{\sqrt{1-e^2}}
             \left(\frac{w}{v_0}\right)^2
             \sin^2\!\psi\sin 2\chi \right] \:, &&
\end{eqnarray}
for the change in the mean anomaly (dropping terms independent of $\chi$)
\begin{eqnarray}\label{eq:deltaM_w}
  \Delta M^{(w)} = -\sqrt{1-e^2}\left[
    \Delta\omega^{(w)} + \Delta\Omega^{(w)}\cos i \frac{}{}\right]
  \hspace{4em} && \nonumber\\ 
    +\frac{\pi m^\ast}{c^2 p}\left[
    12 \hat{\cal Q}_1' eF_e\sqrt{1-e^2}
    \left(\frac{w}{v_0}\right)\sin\psi\sin\chi
  \right. \hspace{4em} && \nonumber\\ \left.
    + \tilde{\cal Q}_2' \left(2F_e\sqrt{1-e^2}-1\right)
    \left(\frac{w}{v_0}\right)^2\sin^2\!\psi\cos 2\chi\right] \:, &&
\end{eqnarray}
where
\begin{equation}\label{eq:q1}
\hat{\cal Q}_1 = \alpha_1^\ast\:(\xi_p-\xi_c) - 2(1-{\cal A}_p)\xi_c^2 
                                              + 2(1-{\cal A}_c)\xi_p^2 \:,
\end{equation}
\begin{equation}\label{eq:q2}
\hat{\cal Q}_2 = \alpha_2^\ast + (1-{\cal A}_c)\xi_p   
                               + (1-{\cal A}_p)\xi_c \:,
\end{equation}
\begin{equation}\label{eq:q1M}
  \hat{\cal Q}_1' = (1 - {\cal A}_p) \xi_c^2 - (1 - {\cal A}_c)\xi_p^2 \:,
\end{equation}
\begin{equation}\label{eq:q2M}
  \hat{\cal Q}_2' = \alpha_2^\ast - 2(1-{\cal A}_c)\xi_p   
                                  - 2(1-{\cal A}_p)\xi_c \:,
\end{equation}
and
\begin{equation}
  F_e = \frac{1}{1+\sqrt{1-e^2}} \:.
\end{equation}
$\psi$ is the angle between $\hat{\bf k}$, the direction of the orbital
angular momentum, and ${\bf w}$. $\chi$ is the angle between the periastron of
the pulsar and the projection of ${\bf w}$ into the orbital plane.

In fully conservative theories of gravity one finds ${\cal A}_p \equiv {\cal
A}_c \equiv 1$ and $\alpha_1^\ast \equiv \alpha_2^\ast \equiv 0$. Consequently
fully conservative theories of gravity do not predict any preferred-frame
effects in the motion of compact binaries, i.e.\ $\hat{\cal Q}_1 = \hat{\cal
Q}_2= \hat{\cal Q}_1' = \hat{\cal Q}_2' = 0$. In GR, in addition to the absence
of preferred-frame effects, one has $\hat{\cal P}=1$.

\subsection{The PPN limit}

In the true post-Newtonian limit for masses with negligible self gravity one 
finds the strong gravity coefficients as functions of the PPN parameters 
$\beta$, $\gamma$, $\alpha_1$, and $\alpha_2$ (Will 1993):
\begin{equation}
  {\cal G} = 1 \:,
\end{equation}
\begin{equation}
  {\cal A}_p = {\cal A}_c = 1 \:,
\end{equation}
\begin{equation}
  {\cal B}^\ast = \mbox{$\frac{1}{3}$}(2\gamma+1) \:,
\end{equation}
\begin{equation}
  {\cal C}^\ast = \mbox{$\frac{1}{7}$}(4\gamma+3+\alpha_1-\alpha_2) \:,
\end{equation}
\begin{equation}
  {\cal D}_p^\ast = {\cal D}_c^\ast = 2\beta - 1 \:,
\end{equation}
\begin{equation}
  {\cal E}^\ast = 1 + \alpha_2 \:,
\end{equation}
and
\begin{equation}
  \alpha_1^\ast = \alpha_1 \:,
\end{equation}
\begin{equation}
  \alpha_2^\ast = \alpha_2 \:.
\end{equation}
Hence
\begin{equation}
  \hat{\cal P} = \mbox{$\frac{1}{3}$} (2 + 2\gamma - \beta) 
               + \mbox{$\frac{1}{6}$} (2\alpha_1 - \alpha_2)\xi_p\xi_c \:,
\end{equation}
\begin{equation}
  \hat{\cal Q}_1 = \alpha_1 (\xi_p-\xi_c) \:,
\end{equation}
\begin{equation}
  \hat{\cal Q}_2 = \alpha_2 \:,
\end{equation}
\begin{equation}
  \hat{\cal Q}_1' = 0 \:.
\end{equation}
\begin{equation}
  \hat{\cal Q}_2' = \alpha_2 \:,
\end{equation}

Nordtvedt and Will~(1972)\nocite{nw72} derive the contribution of a preferred
frame to the advance of periastron and the change in the eccentricity for
planetary orbits (equations 61, 62). In the PPN-limit our
equations~(\ref{eq:deltaomega_w}) and (\ref{eq:deltae_w}) agree with their
results if one assumes $m_p \ll m_c$ and $e \ll 1$. As we restrict our
discussion to semi-conservative theories of gravity, we do not get the
self-acceleration terms related to $\alpha_3$.


\section{Testing preferred-frame effects in binary pulsars}
\label{timing}

We have seen in the previous Section that if a binary pulsar moves with
respect to a preferred frame of reference, changes of its orbital parameters
occur, which should then become apparent in its timing data. Therefore,
assuming a direction and velocity $w$ for the preferred frame\footnote{We do
not make any assumption about the nature of this preferred frame in our
derivation. Only later, we investigate observational data assuming some
natural choices such as the motion of the binary pulsar with respect to the
cosmic microwave background, representing a distinguishable frame of reference
in the Universe (Nordtvedt and Will 1972).} we will demonstrate that timing
data can be used to constrain or even measure the strong field parameters
$\hat{\cal Q}_1$ and $\hat{\cal Q}_2$, which determine the strength of the
preferred-frame effects.

\subsection{Measuring  PFE effects}

Equations (\ref{eq:deltaomega_w}) to (\ref{eq:deltaM_w}) can be converted into
(averaged) first order time derivatives in the orbital elements via
$\langle\dot\omega\rangle = \Delta\omega/P_b$, etc. At a first glance,
comparing these expressions with the observed values for $\dot\omega$, $\dot
x$, and $\dot e$ \footnote{Note that $\dot\Omega$ is not an observable
  quantity.}  obtained in timing observations, seems to provide a method to
measure or at least constrain the strong-field parameters $\hat{\cal Q}_1$ and
$\hat{\cal Q}_2$. In particular, using a measurement for $\dot \omega$ seems
to be very attractive, as this timing parameter is usually the easiest to
determine among the so called ``Post-Keplerian'' parameters. However, this
seemingly straightforward approach of using only the first order time
derivatives has a number of problems: (a) the gravitational mass parameter
$m^\ast$ can usually not be determined in a theory-independent way, (b) the
value of the strong-field parameter $\hat{\cal P}$ is not known, hence
prevents the use of $\langle\dot\omega\rangle$, and (c) for the most promising
test systems the orbit is usually relativistic enough to cause a non-linear
evolution of the orbital parameters. We discuss each these potential problems
in turn.

Without making detailed assumptions on the strong-field properties of the
underlying theory of gravity, most binary pulsars will not allow the
determination of $m^\ast$ and the orbital inclination $i$ as required in
equations (\ref{eq:deltaomega_w}) to (\ref{eq:deltaM_w}). Indeed, most
Post-Keplerian parameters, like the Einstein delay, $\gamma$, or the change in
orbital period due to gravitational wave damping, $\dot P_b$, cannot be used
as they are expected to have strong-field contributions, which are not
quantified within a theory-independent framework as used in this paper.  In
contrast, under some very natural assumptions concerning the leading terms in
the space-time metric in the inter-body region and far from a binary pulsar
system, the observation of a Shapiro delay can be used to determine the sine
of the orbital inclination.  We will discuss this point in more detail later,
when we also demonstrate that under very special circumstances, such
information can be used to actually determine both mass parameters 
$m_p^\ast={\cal G}Gm_p$ and $m_c^\ast={\cal G}Gm_c$.

Even if we could determine $m^\ast$ from post-Keplerian parameters not
involving $\dot\omega$, we still could not extract the preferred-frame
contribution in $\langle\dot\omega\rangle$ as we do not have a 
(theory-independent) value for $\hat{\cal P}$ of equation~(\ref{eq:omegadot1}). 
In other words, usually one cannot separate the preferred-frame terms from the
other relativistic terms.

Moreover, one can precisely measure post-Keplerian parameters only in
relativistic binary pulsars. In such a case, the preferred-frame effects as
described by equations (\ref{eq:deltaomega_w}) to (\ref{eq:deltaM_w}) cannot
be described by a simple linear-in-time expressions, like $e = \langle\dot
e\rangle (t - t_0)$. Fitting for first order time derivatives, therefore,
cannot be used to consistently test for preferred-frame effects in these
systems. This is true for any binary pulsar system where due to a relativistic
advance of periastron the longitude of periastron has changed significantly
since the time of its discovery. For instance, in PSR B1913$+$16 $\omega$ has
advanced by about 135 degrees since its discovery in 1974, and therefore first
order time derivatives cannot be used to constrain preferred frame effects, as
done in Bailey and Kosteleck{\'y}~(2006). The result is a significant change
in $\chi$ since $\chi = const. - \omega(t)$. Consequently, the time derivatives
of the orbital parameters are trigonometric functions of $\omega(t)$ and,
accordingly, the preferred-frame effects impose periodic changes on the
orbital parameters themselves, with frequencies $\dot\omega$ and $2\dot\omega$.

In order to test for preferred-frame effects in such systems, one needs to
include this periodic `signature' of a preferred frame into the timing model,
which is used for fitting the timing data. In the following we will describe a
timing model and apply it to the Double Pulsar.

\subsection{A timing model for the `signature' of preferred-frame effects}

\subsubsection{Case I. $i\simeq90$ degrees}

As outlined in the previous Section, the presence of preferred-frame effects
in the motion of a binary system results in periodic changes of the orbital
elements. In this Section we will extend the existing standard Damour \& Deruelle
(DD) timing model (Damour \& Deruelle 1985, 1986)
to include these
periodic changes and, therefore, allows to fit for the amplitudes of these
preferred-frame effects. We will demonstrate our method by applying it to the
Double Pulsar system, and so we restrict the following discussion to binary
systems with an orbital inclination, $i$, close to 90 degrees. A
generalisation for systems with orbital inclinations significantly less than
90 degrees is straightforward.

\begin{figure}
\centerline{
\psfig{file=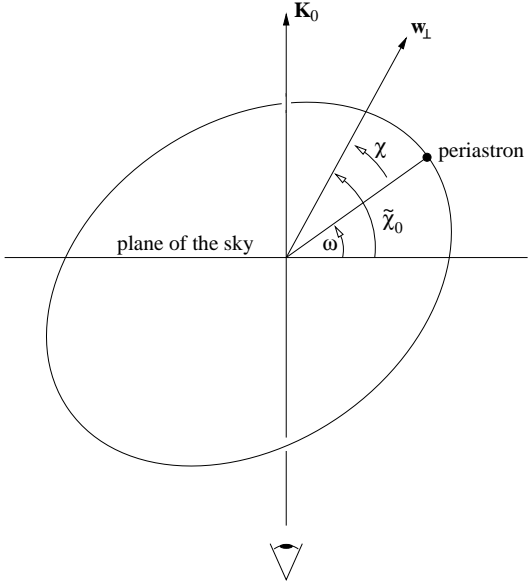,width=7cm}
}
\caption{Definition of angles in the PFE timing model.}
\label{fig:angles}
\end{figure}

If $i \simeq 90^\circ$ changes in the projected semi-major axis (equation
\ref{eq:deltax_w}) are small and can be neglected in the timing model. In the
following $\tilde\chi_0$ denotes the angle between the ascending node and the
projection of ${\bf w}$ into the orbital plane, ${\bf w}_\perp$, and
consequently $\chi=\tilde\chi_0-\omega$ (see Fig.~\ref{fig:angles}).

Integrating equation (\ref{eq:deltaomega_w}) for $i=90^\circ$ one finds
\begin{equation}\label{eq:psromega_w}
  \Delta\omega^{(w)}(T) = \eta_1^{(\omega)} \cos(\omega_L - \tilde\chi_0) 
                        - \eta_2^{(\omega)} \sin 2(\omega_L - \tilde\chi_0) \:,
\end{equation}
where
\begin{equation}\label{eq:psromega_L}
  \omega_L = \omega_0 + \dot\omega^{(1)}(T-T_0) \:,
\end{equation}
and
\begin{equation}\label{eq:eta1q1}
  \eta_1^{(\omega)} = \frac{\hat{\cal Q}_1}{12\hat{\cal P}} 
                      \left(\frac{1}{e}-eF_e^2\right)
                      \left(\frac{w}{v_0}\right)\sin\psi\:,
\end{equation}
\begin{equation}\label{eq:eta2q2}
  \eta_2^{(\omega)} = 
    \frac{\hat{\cal Q}_2}{12\hat{\cal P}} F_e^2
    \left(\frac{w}{v_0}\right)^2\sin^2\!\psi\:.
\end{equation}
When integrating, we have made the assumption that it is sufficient to keep
terms linear in $\eta_i^{(\omega)}$.\footnote{It can be verified directly by
fitting the timing data if this assumption is applicable.} Hence
$\dot\omega^{(1)}$ is given by equation (\ref{eq:omegadot1}).

Integrating equation~(\ref{eq:deltaM_w}) one finds 
\begin{equation}\label{eq:psre_w}
  \Delta e^{(w)}(T) =  \eta_1^{(e)} \sin(\omega_L - \tilde\chi_0) 
                     + \eta_2^{(e)} \cos 2(\omega_L - \tilde\chi_0) \:.
\end{equation}
where
\begin{equation}\label{eq:eta1eq1}
  \eta_1^{(e)} = 
    \frac{\hat{\cal Q}_1}{6\hat{\cal P}} 
    (1+e^2)F_e \left(\frac{w}{v_0}\right)\sin\psi \:,
\end{equation}
\begin{equation}\label{eq:eta2eq2}
  \eta_2^{(e)} = 
    \frac{\hat{\cal Q}_2}{12\hat{\cal P}}
    \frac{e(1+e^2)F_e}{\sqrt{1-e^2}}\left(\frac{w}{v_0}\right)^2\sin^2\!\psi \:.
\end{equation}

Integrating equation~(\ref{eq:deltae_w}) one finds 
\begin{equation}\label{eq:psrM_w}
  \Delta M^{(w)}(T) = -\eta_1^{(M)}\cos(\omega_L - \tilde\chi_0)
                      +\eta_2^{(M)}\sin 2(\omega_L - \tilde\chi_0) \:,
\end{equation}
where
\begin{equation}\label{eq:eta2mq1}
  \eta_1^{(M)} =\eta_1^{(\omega)} \sqrt{1-e^2} +
    \frac{\hat{\cal Q}_1'}{\hat{\cal P}}
    eF_e\sqrt{1-e^2}
    \left(\frac{w}{v_0}\right)^2\!\sin^2\!\psi \:,
\end{equation}
\begin{eqnarray}\label{eq:eta2mq2}
  \eta_2^{(M)} = \eta_2^{(\omega)} \sqrt{1-e^2} +
    \frac{\hat{\cal Q}_2'}{6\hat{\cal P}}
    \left(F_e\sqrt{1-e^2} - \mbox{$\frac{1}{2}$}\right)
    \left(\frac{w}{v_0}\right)^2\!\sin^2\!\psi \:.
  \hspace{-2em} && \nonumber\\ && 
\end{eqnarray}
We emphasize that the ratios $\eta_1^{(\omega)}/\eta_1^{(e)}$ and
$\eta_2^{(\omega)}/\eta_2^{(e)}$ are only functions of the Keplerian
eccentricity, $e$, of the binary system.

For a consistent, theory-independent analysis of binary pulsar timing data,
Damour and Deruelle developed a phenomenological timing model, using a
parametrized post-Keplerian approach (Damour \& Deruelle 1985,
1986\nocite{dd85,dd86}). The DD timing model applies to the class of
Lorentz-invariant theories of gravity, such as the well-motivated
tensor-scalar theories. However, it does not describe the preferred-frame
effects that we consider here. On the other hand, it is easy to incorporate
these variations into the DD timing model by simply adding solutions
(\ref{eq:psromega_w}), (\ref{eq:psre_w}), (\ref{eq:psrM_w}) accordingly to the
the DD model: 
\begin{equation}
\omega(T) \rightarrow \omega(T) + \Delta \omega(T)^{(w)} \:, \\
\end{equation}
\begin{equation}
e(T) \rightarrow e(T) + \Delta e(T)^{(w)} \:, \\
\end{equation}
\begin{equation}
\frac{2\pi}{P_b} (T-T_0) \rightarrow \frac{2\pi}{P_b}(T-T_0) 
                                       + \Delta M(T)^{(w)} \:.
\end{equation}
Therefore our new timing formula, for binary pulsars with $i \simeq 90^\circ$,
contains, in addition to the Keplerian and post-Keplerian parameters of the DD
timing model, five new preferred-frame timing parameters: the four PFE
amplitudes $\eta_1^{(\omega)}$, $\eta_2^{(\omega)}$, $\eta_1^{(M)}$,
$\eta_2^{(M)}$, and the angle $\tilde\chi_0$. The amplitudes $\eta_1^{(e)}$
and $\eta_2^{(e)}$ are proportional to $\eta_1^{(\omega)}$ and
$\eta_2^{(\omega)}$ respectively, where the factors are only functions of the
Keplerian eccentricity, $e$.

If the longitude of periastron has advanced for a considerable amount ($\ga
90^\circ$) since the discovery of a binary pulsar, the timing data will allow
to fit for all of these new timing parameters.  If the binary motion shows the
effect of a preferred-frame in the Universe, all five independent
`preferred-frame parameters', $\eta_1^{(\omega)}$, $\eta_2^{(\omega)}$,
$\eta_1^{(M)}$, $\eta_2^{(M)}$, and $\tilde\chi_0$ can be determined. One can
even decide to fit additionally for $\eta_1^{(e)}$ and $\eta_2^{(e)}$
i.e.~perform a fit simultaneously for these parameters and for
$\eta_1^{(\omega)}$ and $\eta_2^{(\omega)}$. The existence of a preferred
frame could then be confirmed by verifying the expected relationship between
the PFE amplitudes in $\omega$ and $e$.  The ratios of these amplitudes are
unique in the sense that they are a function of $e$ only and therefore
independent of any strong-field parameters, and of the magnitude and direction
of ${\bf w}$. For the parameters of the Double Pulsar we would expect to measure
\begin{equation}
  \eta_1^{(\omega)} / \eta_1^{(e)} = 11.262 \:,
\end{equation}
\begin{equation}
  \eta_2^{(\omega)} / \eta_2^{(e)} = 5.642 \:
\end{equation}
if a preferred frame exists.
This first test would incontrovertibly reveal the existence of a preferred frame.
We will show later, how the direction of the preferred can also be constrained
or even determined.

In the absence of any preferred-frame effect, we expect none of the PFE timing
parameters to have significant values. In particular, the magnitude of PFE
amplitudes should be consistent with zero, and the angle $\tilde\chi_0$ is
undefined. In this case, one can obtain upper and lower limits for
$\eta_1^{(\omega)}$, $\eta_2^{(\omega)}$, $\eta_1^{(M)}$, and $\eta_2^{(M)}$
by holding $\tilde\chi_0$ fixed to a particular value while fitting for the
PFE amplitudes. Stepping through the possible values $\tilde\chi_0\in \left[0,
2\pi\right]$, various directions on the sky can be probed. In the case of a
relativistic binary pulsar, a clear separation of the preferred-frame
contribution ($\eta_1^{(\omega)}$, $\eta_2^{(\omega)}$) from the other
relativistic terms ($\dot\omega^{(1)}$) in the precession of periastron is
possible and will improve with time due to increasing coverage of
$\omega$-space.

\subsubsection{Case II. $i\ne90$ degrees}

As stated earlier, the above assumes that $i \simeq 90^\circ$. If one allows
for any orbital inclinations, one finds that equation~(\ref{eq:psromega_w}) is
replaced by
\begin{equation}\label{eq:psromega2}
  \Delta\omega =
      \eta_1^{(\omega)'} \cos(\omega_L - \tilde\chi_0 + \delta_1') 
    - \eta_2^{(\omega)'} \sin 2(\omega_L - \tilde\chi_0 + \delta_2') \:.
\end{equation}
Like $\eta_1^{(\omega)}$ and $\eta_2^{(\omega)}$, $\eta_1^{(\omega)'}$ and
$\eta_2^{(\omega)'}$ are proportional to $\hat{\cal Q}_1/\hat{\cal P}$ and
$\hat{\cal Q}_2/\hat{\cal P}$ respectively, while $\delta_1'$ and $\delta_2'$
depend only on Keplerian parameters and the orientation of the pulsar with
respect to the preferred frame. In this case also $x(T)$ has to be considered,
for which one finds an expression similar to equation~(\ref{eq:psromega2}). 
Following the procedure above, a derivation
of these expressions is straight forward.

\subsection{Translating timing model parameters to strong-field PFE parameters}

In this Section we give a description on how to convert the timing results,
obtained after applying the new {\em PFE timing model}, into the strong-field
PFE parameters introduced in Section~\ref{motion}.

As discussed in the previous Section, for a binary pulsar with $i \simeq
90^\circ$ values for the timing parameters $\eta_1^{(\omega)}$,
$\eta_2^{(\omega)}$, $\eta_1^{(M)}$, and $\eta_2^{(M)}$ can be converted into
values for $\hat{\cal Q}_1 w \sin\psi / \hat{\cal P}$, $\hat{\cal Q}_2
w^2\sin^2\psi / \hat{\cal P}$, $\hat{\cal Q}_1' w \sin\psi / \hat{\cal P}$,
$\hat{\cal Q}_2' w^2\sin^2\psi / \hat{\cal P}$.  Without any further
assumptions, however, we cannot determine or even restrict the angle $\psi$.
Here one has to keep in mind that, in general, the longitude of the ascending
node of the pulsar orbit, $\Omega$, cannot be determined from pulsar
timing observations and has to be treated as a free parameter. We discuss
possible exceptions further below.

Relating the orientation of the orbit to the movement relative to a preferred
frame in direction ${\bf w}$, requires the knowledge of the orbital
inclination angle. In cases where the binary orbit is seen nearly edge-on, one
can often determine this angle accurately, as one expects to observe an
additional delay in the arrival time of the pulsar signal during superior
conjunction of the pulsar, caused by the gravitational field of the companion.
This is the case in the Double Pulsar where this ``Shapiro effect'' can be
used to determine the orbital inclination $i$ (modulo the ambiguity $i
\rightarrow \pi-i$). In order to do this, one only have to make some
general, very natural assumptions. As in Will~(1993), we assume that to
first order the metric valid in the inter-body region and far from the system
can be written as
\begin{eqnarray}
g_{00} &=& -1 + \frac{2G\kappa_p^\ast m_p}{c^2|{\bf x}-{\bf x}_p|}
              + \frac{2G\kappa_c^\ast m_c}{c^2|{\bf x}-{\bf x}_c|} 
              \label{eq:g00} \\
g_{0j} &=& 0  \label{eq:g0j} \\
g_{ij} &=& \left(1 + \frac{2G\gamma_p^\ast m_p}{c^2|{\bf x}-{\bf x}_p|}
                   + \frac{2G\gamma_c^\ast m_c}{c^2|{\bf x}-{\bf x}_c|}
           \right)\delta_{ij} \label{eq:gij}
\end{eqnarray}
where $\kappa_p^\ast$, $\gamma_p^\ast$, $\kappa_c^\ast$, and $\gamma_c^\ast$
are functions of the parameters of the theory and of the structure of pulsar
and companion. In GR $\kappa_p^\ast = \gamma_p^\ast = \kappa_c^\ast =
\gamma_c^\ast = 1$. The time taken by the pulsar signal to travel from the
pulsar to the solar system in such a metric can be accounted for by adding
\begin{eqnarray}
  \Delta_S = 
    \frac{G(\kappa_c^\ast+\gamma_c^\ast)m_c}{c^3}\ln\left\{1-e\cos U \frac{}{}
    \right. \hspace{7em} && \nonumber\\ \left. 
    - \sin i \left[\sin\omega(\cos U - e) \frac{}{}
    \right.\right.\hspace{3em} && \nonumber\\ \left.\left. \frac{}{} 
    + (1-e^2)^{1/2}\cos\omega\sin U\right]\right\}\:. &&
  \label{eq:shap}
\end{eqnarray}
to the timing model, which can be used to fit for $\sin i$, the ``shape'' of
the Shapiro delay, without any knowledge of the strong-field parameters
(Blandford and Teukolsky~1976\nocite{bt76}, Damour and Deruelle~1986b,
Will~1993). $U$ is the eccentric anomaly as defined by the DD model. The
measurement of $\sin i$ then allows to exclude certain directions of ${\bf w}$
in the sky, given certain values of $\tilde\chi_0$. In fact, if one makes
certain assumptions of the magnitude and direction of ${\bf w}$, i.e.\ the
direction and motion of the preferred frame, a measurement of $\sin i$ (and
the computation of $\tilde\chi_0$) will allow the determination of $\sin\psi$,
and, consequently, the determination of $\hat{\cal Q}_1 / \hat{\cal P}$ and
$\hat{\cal Q}_1 / \hat{\cal P}$ for the assumed preferred frame from the
fitted PFE amplitudes.

On the other hand, as outlined earlier, if no preferred-frame effects are
observable in the orbital motion of a binary pulsar, fitting for the PFE
amplitudes (while holding $\tilde\chi_0$ fixed) can be used to determine
limits on $\hat{\cal Q}_1 w/ \hat{\cal P}$, $\hat{\cal Q}_2 w^2/ \hat{\cal
P}$, $\hat{\cal Q}_1' w/ \hat{\cal P}$, and $\hat{\cal Q}_2' w^2/ \hat{\cal
P}$ for (nearly) any direction in the sky. However, during this transformation
from PFE timing model parameters to these physical quantities, one needs to
determine $\tilde\chi_0$ and $\psi$ for a given direction in the sky. That
requires the knowledge of the, in general, unknown angle $\Omega$, while one
also has to account for the ambiguity in the sense of the inclination $i$. Two
practical approaches exist to overcome this problem.  These angles can either
be chosen such that the limits are most conservative, or one can perform
Monte-Carlo simulations varying over $\Omega$ and the two possible values for
$i$. Assuming a uniform distribution for $\Omega$ and equal probability for
$i$ and $\pi-i$, it is possible to determine limits for any chosen confidence
level. It is clear that the most conservative method cannot provide any
restrictions for directions along a cone that has an opening angle $i$ around
the line of sight, as for any direction on this cone, $\Omega$ can be chosen
such that $\psi=0$.

The situation is obviously improved if $\Omega$ can be determined.
We note that the proper motion of a binary pulsar can produce secular changes
in the orbital elements that depend on the longitude of the ascending node,
$\Omega$, and the orbital inclination, $i$, as it changes the
apparent geometrical orientation of the orbit (Arzoumanian et
al.~1996\nocite{ajrt96} and Kopeikin~1996\nocite{kop96}). In principle these
changes can be used to determine $\Omega$ and $i$ from timing
observations. However, in practice it will be difficult to separate the proper
motion effects from the relativistic changes of the binary orbit, in
particular if we make only very generic assumptions about the underlying
theory of gravity, as we do in this paper.

\label{iss}
In principle, measurements of the timescale of the interstellar scintillation
(ISS) over an orbit can also be used to estimate the orbital inclination $i$ and
the transverse velocity of the centre of mass of the system (Lyne \&
Smith~1982\nocite{ls82}, Ransom et al.~2004\nocite{rkr+04}, Coles et
al.~2005\nocite{cmr+05}). A comparison of the transverse velocity derived from
ISS and the transverse velocity obtained from timing observation could then
be used to determine $\Omega$. However, the scintillation-based
velocity depends on a number of assumptions about the properties of the
effective scattering screen and, therefore, are more susceptible to systematic
errors than timing measurements.


\section{The Double Pulsar}
\label{psr0737}

Following the discovery of the Double Pulsar system in April 2003
(Burgay~et~al.~2003\nocite{bdp+03}; Lyne~et~al.~2004\nocite{lbk+04}), timing
observations have enabled the most stringent tests of GR in the strong-field
regime to date (Kramer et al.~2006)\nocite{ksm+06}. Indeed, the system is a
unique laboratory for gravitational physics for a number of reasons. Firstly,
both members of the binary system are visible as active radio pulsars,
providing access to a simple measurement of the ratio of their masses and
hence providing theory-independent (at least to 1PN order) constraints for
tests of theories of gravity.

Secondly, the fortunate orientation of the binary orbit in space allows us to
observe the Double Pulsar under a nearly perfect edge-on geometry. In addition
to precise measurements of a Shapiro delay, this enables further independent
estimates of the system's inclination angle. 
\footnote{Observations and modelling of the
eclipse of the radio emission of the millisecond pulsar PSR~J0737$-$3039A
during superior conjunction, lasting for about 25-30 seconds, provides another
estimate (Breton~et~al.~2006\nocite{bkm+06}), while two methods using observed
scintillation properties provide two others
(Ransom~et~al.~2004\nocite{rkr+04}; Coles~et~al.~2005\nocite{cmr+05}).}

The combination of the determination of $\sin i$ via the Shapiro delay with
the measurement of the mass ratio allows a theory-independent determination of
both mass parameters, $m_p^\ast$ and $m_c^\ast$, as we will discuss in detail
later. In this case $\hat{\cal P}$ can now be measured independently, thus
directly giving limits for $\hat{\cal Q}_1$ and $\hat{\cal Q}_2$ from fitting
for $\eta_1^{(\omega)}$ and $\eta_2^{(\omega)}$.  Further, from this one can
determine $\alpha_1^\ast$ and $\alpha_2^\ast$ for gravity theories that fulfil
${\cal A}_p \equiv {\cal A}_c \equiv 1$. In this sense the Double Pulsar is
indeed a unique laboratory for strong-field preferred-frame effects.

The third reason why the Double Pulsar is a unique gravity lab is the
compactness of the orbit. With larger mean orbital velocities than in any
other binary pulsar, the system is the most relativistic system known. This is
reflected by the measurement of the largest advance of periastron observed in
any binary system. As a result, the Double Pulsar orbit precesses by 16.9
degrees per year with respect to any possible existing preferred frame,
changing the angle $\chi$ in Fig.~\ref{fig:angles} by a corresponding
amount. As such the Double Pulsar is an ideal candidate to study the existence
of preferred-frame effects in strong-field gravity as the magnitude of
$\omega$ does not only allow its very precise measurement (currently at the
$4\times 10^{-5}$ level) but the orbit also `covers' a lot of periastron angle
space in a rather short time span. According to the theoretical framework laid
out in the previous sections, such a system is ideal for testing for PFE
variations in the orbital parameters, the `signature' of a preferred frame.

\begin{figure}
\centerline{
\psfig{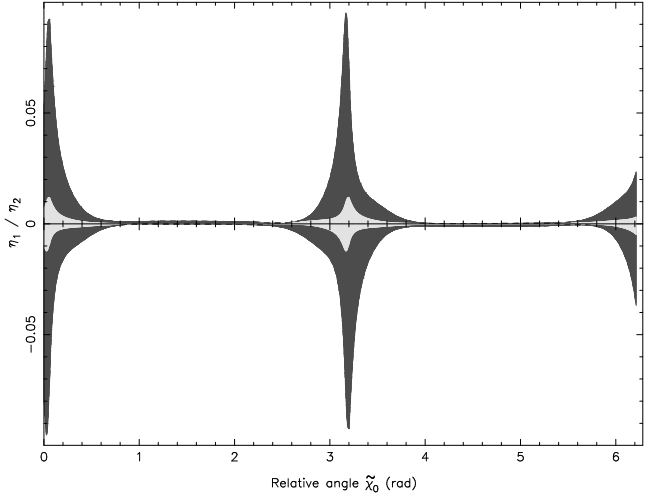}
}
\caption{\label{fig:eta1eta2} Limits on the existence of preferred-frame
effects present in the orbital motion of the Double Pulsar as measured by the
parameters $\eta_1^{(\omega)}$ (dark grey) and $\eta_2^{(\omega)}$ (light
grey) as introduced in the PFE timing model. See text for details.}
\end{figure}

\subsection{Application of the PFE timing model}

In Section~\ref{timing}, we have developed the {\em PFE timing model} that
introduces, based on the DD timing model, the new PFE model parameters
$\eta_1^{(\omega)}$, $\eta_1^{(\omega)}$, $\eta_1^{(M)}$, $\eta_2^{(M)}$
$\tilde\chi_0$. We now want to apply this model to the timing data of the
Double Pulsar. We make use of nearly three years of timing data obtained by
Kramer et al.~(2006) during which the pulsar periastron has advanced by about
50 degrees.

At present the timing data of the Double Pulsar system does not show a PFE
`signature'. Hence, we can use the data only to derive limits for the PFE
amplitudes by fitting for the PFE amplitudes while holding $\tilde\chi_0$
fixed at a given value. We then vary $\tilde\chi_0$ in sufficiently small
steps between $0$ and $360^\circ$ to derive limits for any value of
$\tilde\chi_0$. Fig.~\ref{fig:eta1eta2} illustrates the 95\% confidence for
$\eta_1^{(\omega)}$ and $\eta_2^{(\omega)}$, consistent with the non-existence
of preferred frames.

In principle, we would also aim to fit for $\eta_1^{(M)}$ and $\eta_2^{(M)}$
which would yield limits for $\hat{\cal Q}_1'$ and $\hat{\cal Q}_2'$, giving
separate limits for $\alpha_1^\ast$, $\alpha_2^\ast$, ${\cal A}_p$, ${\cal
  A}_c$ when combined with the results for $\eta_1^{(\omega)}$ and
$\eta_2^{(\omega)}$.  However, the Double Pulsar timing data used in this
paper do not yet allow to fit for $\eta_1^{(M)}$ and $\eta_2^{(M)}$. The
periastron has not advanced far enough to separate these PFE amplitudes from
other orbital parameters, in particular the decrease in the orbital period due
to the emission of gravitational waves.  As we will show later, with longer
time-spans and more coverage of the periastron angle, such a fit will be
possible, while simultaneously the amplitude of all limits will be greatly
reduced. It is obvious that some orientations of the orbit with respect to
{\bf w} lead to better constraints than others.

\begin{figure*}
\centerline{
\begin{tabular}{cc}
\psfig{file=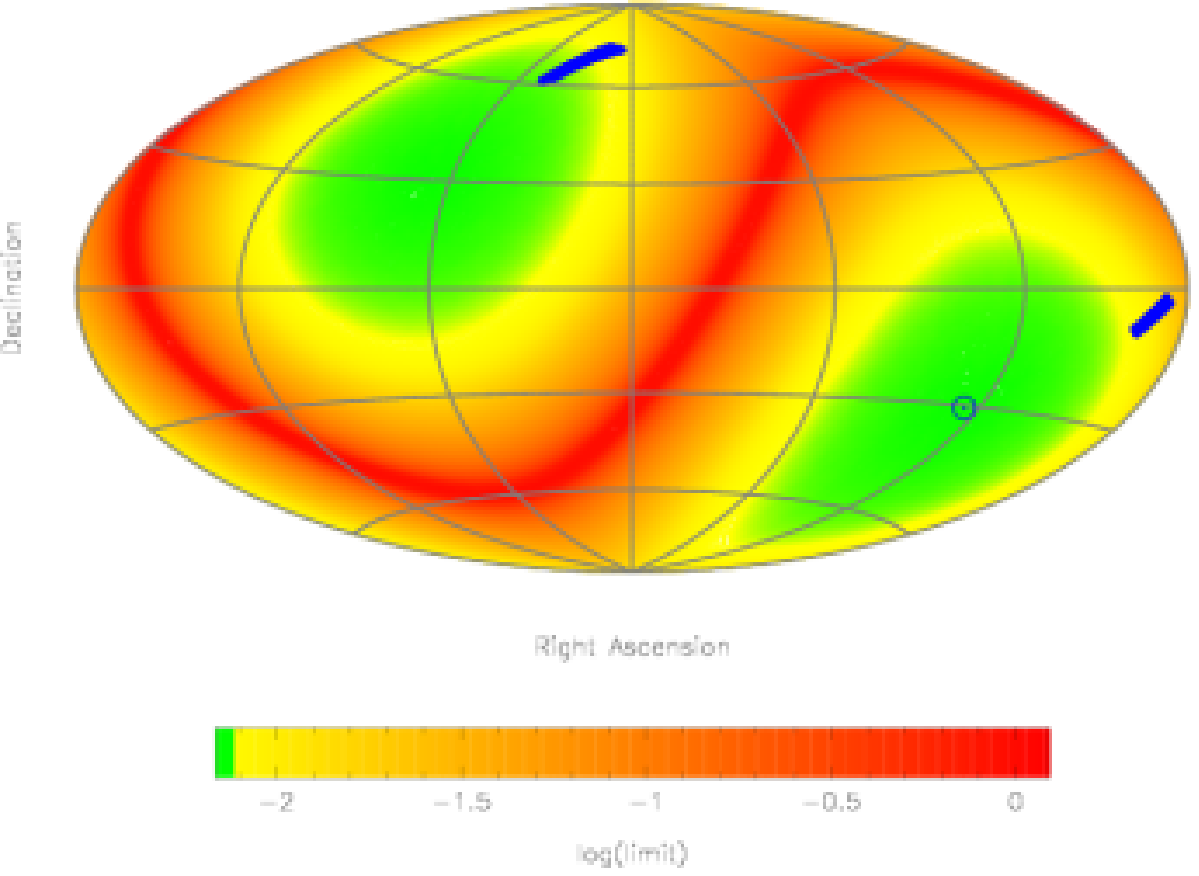,width=8.5cm,angle=-90} &
\psfig{file=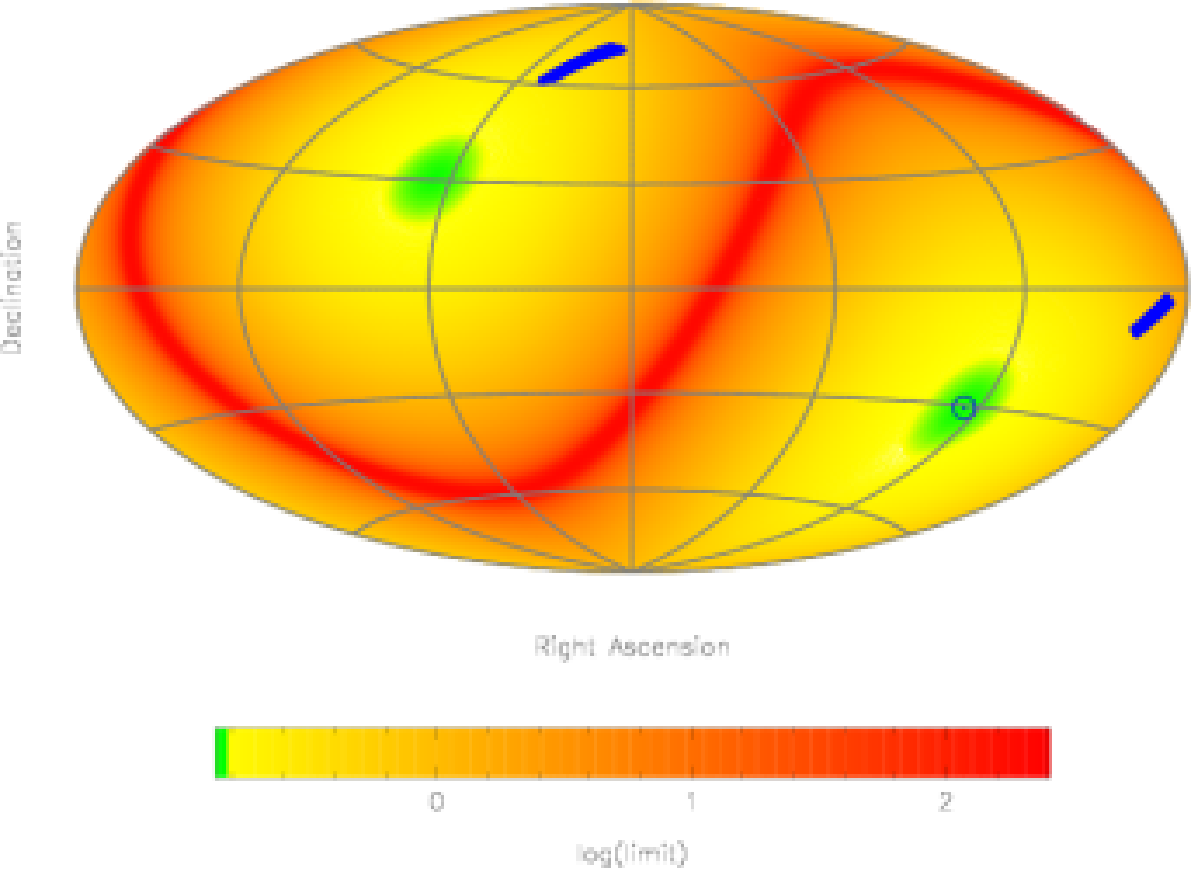,width=8.5cm,angle=-90} 
\end{tabular}
}

\caption{\label{fig:q1}\label{fig:q2}
Limits computed for 95\% confidence levels on $|{\cal Q}_1|$ and $|{\cal
   Q}_2|$ for the sky as seen by the Double Pulsar. The limits are calculated
for $|{\bf w}| = 100$ km/s, and have to be scaled accordingly for other
values of $|{\bf w}|$. Directions discussed in the text are marked by the
blue area, i.e.~direction relative to a Galactic frame (top left) and
direction relative to the Cosmic Microwave Background (CMB) (right). The
extension of these areas reflects the uncetainty in the radial velocity of
the binary pulsar system with respect to the solar system. When extracting
the limits for the Galactic and the CMB frame from these figures, one needs
to keep in mind that the $|{\bf w}|$ of these frames changes across these
areas. The location of the Sun as seen from the Double Pulsar is marked by
$\odot$.}

\end{figure*}

\subsection{Determination of quantitative limits}

In this Section we translate the limits on the `signature' of a preferred
frame into quantitative limits within the generalised EIH formalism. Our goal
is, first, to determine limits for $\hat{\cal Q}_1w$ and $\hat{\cal Q}_2w^2$
for any direction in the sky, and secondly, to determine limits for $\hat{\cal
  Q}_1$ and $\hat{\cal Q}_2$ for a preferred frame that is at rest with
respect to the CMB and one that is at rest with respect to our galaxy.

The first step is to determine the masses from the timing data described by
applying the PFE timing model. Equation~(\ref{eq:shap}) can be used to convert
the measurement of the Shapiro ``shape'' timing parameter into a measurement
of $\sin i$. Now, combining equation~(3.15) of Damour and
Taylor~(1992)\footnote{Note that the effective gravitational constant ${\cal G}$
in Damour and Taylor~(1992) corresponds to ${\cal G}G$ in this paper.} with the
measurement of the mass ratio obtained from $R$ in the Double Pulsar, we obtain
the following values
\begin{equation}
   m_p^\ast = {\cal G}Gm_p = (1.339 \pm 0.002) \times GM_\odot\:,
\end{equation}
\begin{equation}
   m_c^\ast = {\cal G}Gm_c = (1.250 \pm 0.002) \times GM_\odot \:.
\end{equation}

It is at this point where we exploit the uniqueness of the Double Pulsar which
provides us with measurements of $m_p^\ast$ and $m_c^\ast$. Combining these
with the observed rate of periastron advance allows us to determine the
strong-field parameter $\hat{\cal P}$ (see equation~\ref{eq:omegadot1}). The
result is
\begin{equation}\label{eq:numphat}
  \hat{\cal P} = 1.000 \pm 0.001 \:.
\end{equation}

In order to establish the absolute orientation of the binary system in space,
one needs the orientation of the ascending node $\Omega$, which is still
unknown for the Double Pulsar (but see discussion in Section~\ref{timing}).
As described earlier, we perform Monte-Carlo simulations and vary this
parameter between 0 and $2\pi$ for any given direction of ${\bf w}$. For a
given ${\bf w}$ and a given $\Omega$, we compute the angle $\tilde\chi_0$ and
read off the corresponding limits for $\eta_1^{(\omega)}$ and
$\eta_2^{(\omega)}$ from the calculated and tabulated values shown in
Fig.~\ref{fig:eta1eta2}. To account also for the ambiguity in the sense of the
inclination $i$, the whole procedure is performed for 10,000 datasets for each
direction in the sky and inclination angles of $i$ and $\pi-i$.  As discussed
earlier, the obtained values for $\eta_1^{(\omega)}$ and $\eta_2^{(\omega)}$
can be translated into limits for $\hat{\cal Q}_1w$ and $\hat{\cal Q}_2w^2$
using equations~(\ref{eq:eta1q1}) and (\ref{eq:eta2q2}). Figs.~\ref{fig:q1}
give the results of our numerical simulations for preferred frames associated with
different directions of the sky as seen from the Double Pulsar. The limits
shown as a colour map are derived for an assumed binary pulsar velocity
relative to the preferred frame of 100~km/s. Results for other relative speeds
can be scaled according to equations~(\ref{eq:eta1q1}) and (\ref{eq:eta2q2}).

\begin{figure*}
\centerline{
\begin{tabular}{cc}
\psfig{file=fig4a.ps,width=8.cm,angle=-90} &
\psfig{file=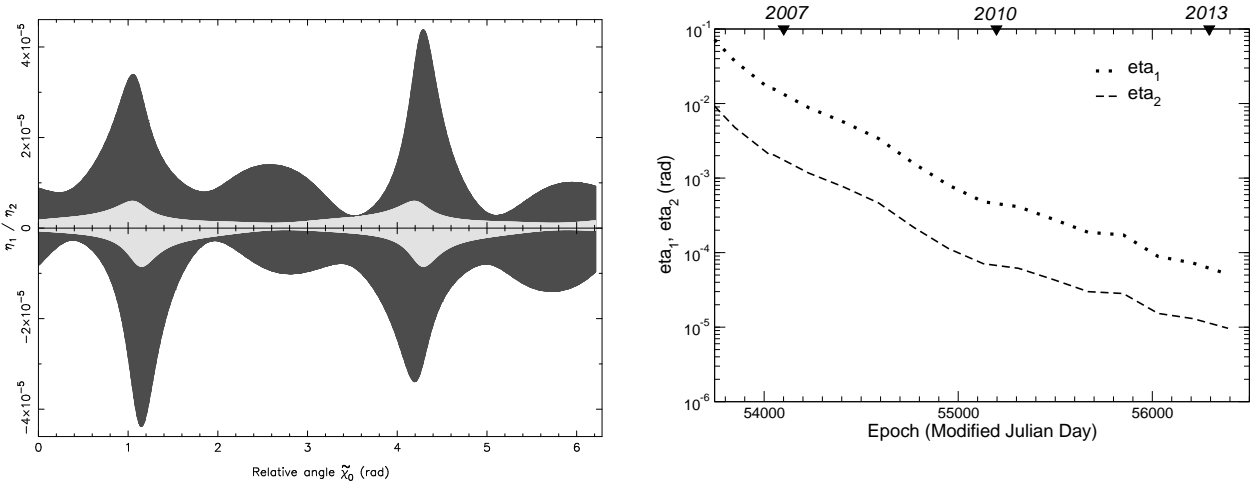,width=9.2cm,angle=-90} 
\end{tabular}
}

\caption{ Future improvement of PFE tests in the Double Pulsar.
(Left) Amplitude of the PFE 
parameters $\eta_1^{(\omega)}$ (dark grey) and $\eta_2^{(\omega)}$ (light
grey) as expected for 10 years of Double Pulsar timing data.
(Right) Evolution of limits on $\eta_1^{(\omega)}$ and
$\eta_2^{(\omega)}$ as a function of time by recording their worst values
over all $\tilde\chi_0$ at a given epoch. The year of observation is 
indicated at the top of the figure.
\label{fig:etafuture} }
\end{figure*}

We also indicate the directions for two selected frames that may be considered
as being related to a preferred frame; we indicate the direction of motion
relative to the CMB and the direction of the motion relative to the Galactic
reference frame. The motion of the solar system with respect to the CMB we
take from Hinshaw et al.~(2006)\nocite{hnb+06} and Lineweaver et
al.~(2006)\nocite{lts+96}, and with respect to the Galactic reference frame
from Mignard~(2000)\nocite{mig00}. To calculate the motion of the Double
Pulsar with respect to these frames of reference we need to know the motion of
the Double Pulsar with respect to the solar system. While a proper motion of
the pulsar has been measured, one of the few unknown system parameters is the
undetermined radial velocity of the system. In accordance with recent
evolution studies of the Double Pulsar system
(Stairs~et~al.~2006\nocite{std+06}), we perform our analysis for a range of
radial velocities between $-100$ and $+100$~km/s, and therefore, the direction
relative to the CMB and the Galactic frame convert into small areas on the
sky, from which we pick the worst 95\% C.L. limits on $\hat{\cal Q}_1$ and
$\hat{\cal Q}_2$:
\begin{eqnarray}
  \mbox{CMB:} & -0.02 < {\cal Q}_1 < 0.01 & -0.3 < {\cal Q}_2 < 0.2 \\
  \mbox{GAL:} & -0.01 < {\cal Q}_1 < 0.01 & -0.3 < {\cal Q}_2 < 0.3 
\end{eqnarray}
For gravity theories with ${\cal A}_p \equiv {\cal A}_c \equiv 1$ we obtain
directly limits for $\alpha_1^\ast$ and $\alpha_2^\ast$ (see
equations~\ref{eq:q1} and \ref{eq:q2}):
\begin{eqnarray}
  \mbox{CMB:} & -0.5 < \alpha_1^\ast < 0.3 & -0.3 < \alpha_2^\ast < 0.2 \\
  \mbox{GAL:} & -0.3 < \alpha_1^\ast < 0.3 & -0.3 < \alpha_2^\ast < 0.3 
\end{eqnarray}

The limits given here are not very tight yet. In fact, they are several orders
of magnitude worse than the limits for the weak-field PPN parameters
$\alpha_1$ and $\alpha_2$. On one hand, as we will discuss below, these
limits will improve considerably during the next couple of years. On the other
hand, these limits hold for preferred-frame effects related to strong
gravitational fields. If one expands, for instance, $\alpha_1^\ast$ as a
function of the sensitivity of the gravitation body
\begin{equation}
\label{eqn:al1}
  \alpha_1^\ast = \alpha_1 + \alpha_{11}(s_p + s_c) + \dots
\end{equation}
one sees that tests in the solar system made to restrict $\alpha_1$, can
only test terms which are not related to the internal gravitational fields of
a body. Testing these strong-field coefficients, like $\alpha_{11}$, has to be
done via binary pulsars.

Our strong-field limit for $\alpha_1^\ast$ has to be compared with the limits
obtained by Damour and Esposito-Far{\`e}se~(1992b) and Wex~(2000), which are
better by about three orders of magnitude.  However, in a few years from now
we expect the limits obtained from the Double Pulsar to be comparable with the
limits obtained from small-eccentricity binary pulsars, and as outlined in
Section~\ref{timing} the method presented here does not need to rely on
probabilistic considerations.

Emphasizing the arguments and results presented by Will (1993) and Damour and
Esposito-Far\`ese (1996a), we point out that Double Neutron star Systems (DNSs)
are particularly sensitive to theories of gravity where higher order terms
in Eqn.~(\ref{eqn:al1}) become important, so that for instance
$\alpha_1^\ast \propto s_p s_c$ (similarly, 
$\alpha_2^\ast \propto s_p s_c$), since for a white dwarf 
$s_{\rm WD}\la 10^{-3} \ll s_{\rm NS}$.


\subsection{Future measurements}
\label{future}

The continuing precession of periastron in the Double Pulsar will lead
to a constant improvement in the limits on preferred frames using our
proposed method. In order to predict the future limits we have
simulated timing data for the Double Pulsar, assuming similar timing
precision and frequencies of observations as described in
Kramer~et~al.~(2006). Simulating data until 2013, hence for 10 years
after the discovery of the system and half of the periastron
precession period, we have computed the limits on the observed
parameters $\eta_1^{(\omega)}$ and $\eta_2^{(\omega)}$ as a function
of time. In these simulations we make the unlikely and hence
conservative assumption that the timing precision will fail to improve
over the next years, so that the quoted limits should be considered as
conservative also. The result is shown in
Fig.~\ref{fig:etafuture}. Both parameters scale essentially similarly,
and in 2013 the limits will have improved by more than two orders of
magnitude, resulting in limits of $|\alpha_1^\ast| < 2 \times 10^{-3}$
and $|\alpha_2^\ast| < 2 \times 10^{-3}$. The vast improvement is also
demonstrated by comparing Figure 2 with the left part of Figure
\ref{fig:etafuture} where we plot the limits on $\eta_1^{(\omega)}$
and $\eta_2^{(\omega)}$ as a function of $\tilde\chi_0$ as expected
for the year 2013. These predicted values are still somewhat worse
than those derived by Damour and Esposito-Far{\`e}se~(1992b) and
Wex~(2000), but we estimate that with the timing precision being
certain to improve over the next few years (e.g.~by using the
telescopes to be constructed as pathfinders to the Square-Kilometre
Array (SKA)), the limits will be improved upon further.

\begin{figure*}
\centerline{
\begin{tabular}{ccc}
\psfig{file=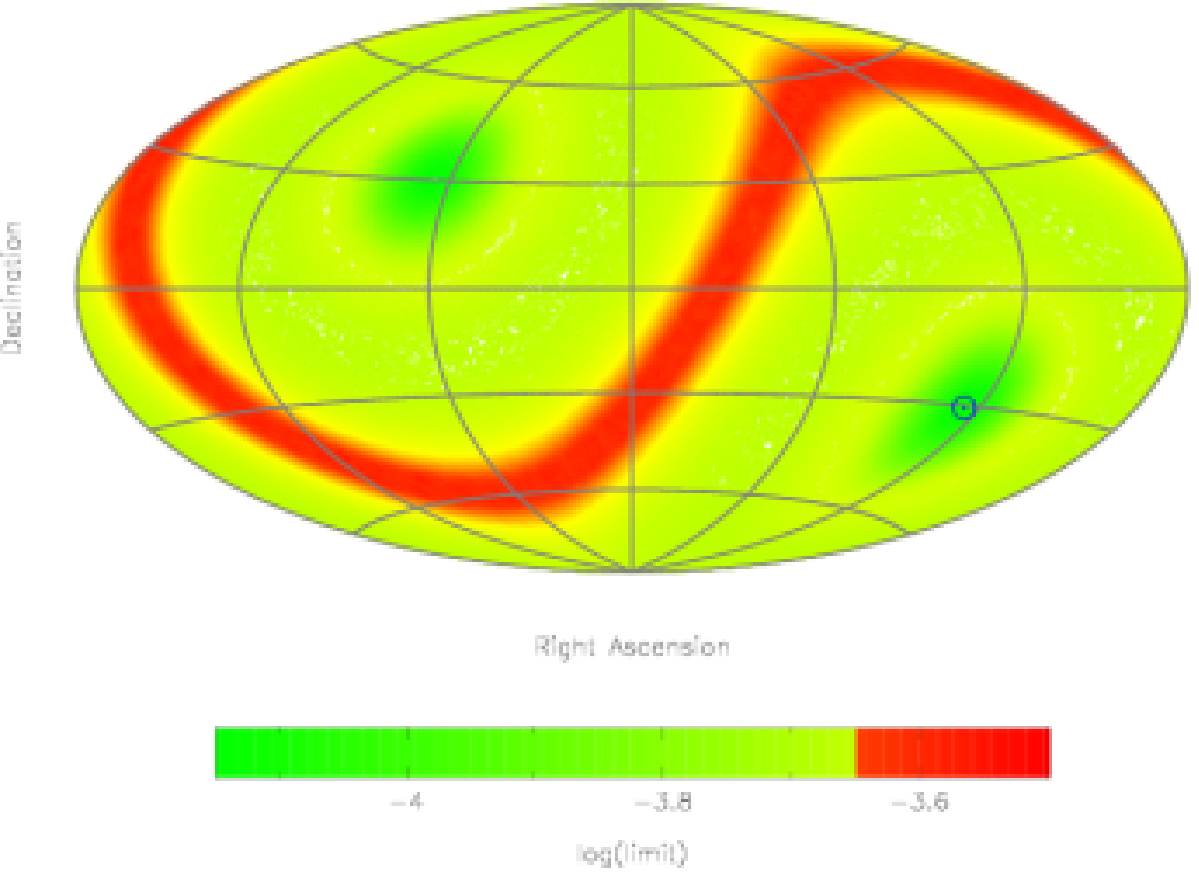,width=6cm,angle=-90} &
\psfig{file=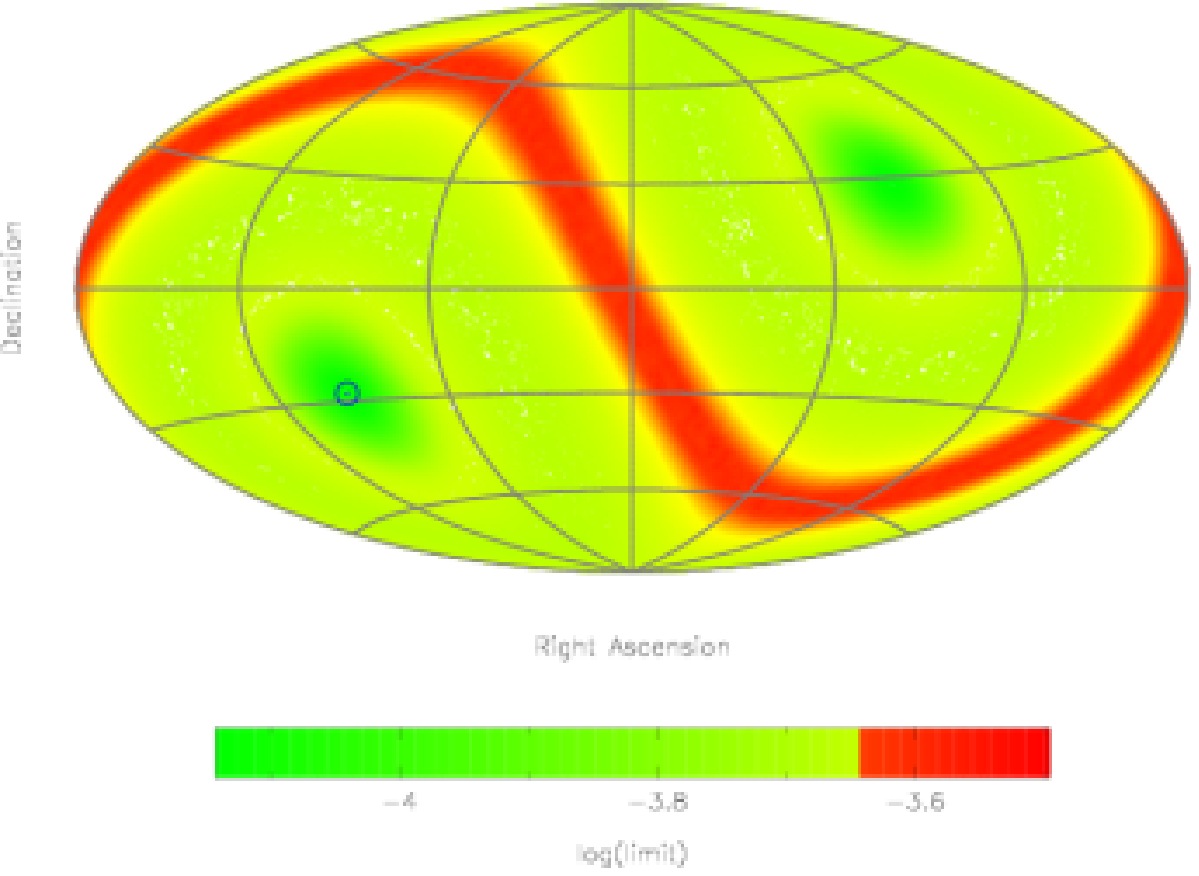,width=6cm,angle=-90} &
\psfig{file=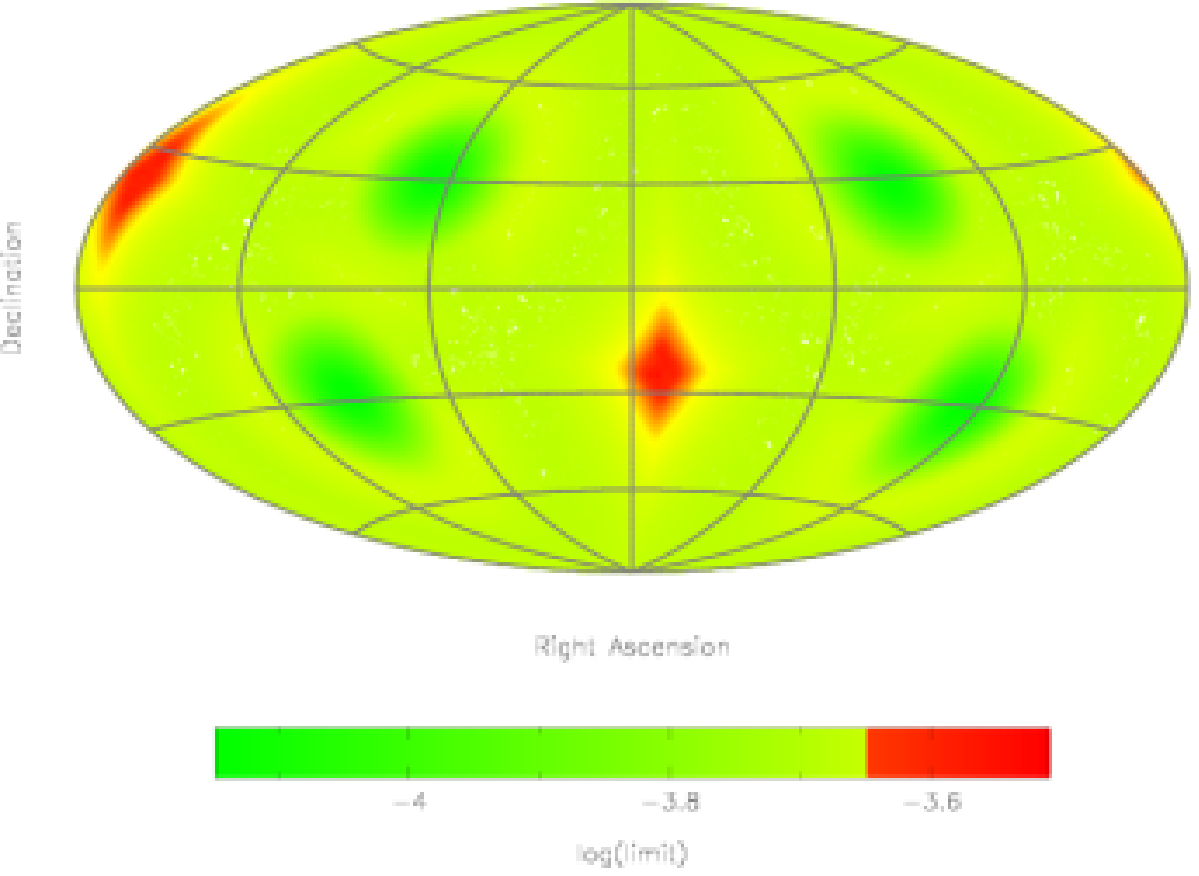,width=6cm,angle=-90} 
\end{tabular}
}

\caption{
  Limits computed for 95\% confidence limits on $|{\cal Q}_1|$ and 
  $|{\cal Q}_2|$ for the sky as seen by the Double Pulsar in 2013 (left) and a
  simulated Double Pulsar-like system located in the Galactic Centre (middle).
  Combining the independent constraints provided by the two systems (right)
  makes this PFE antenna array sensitive to most directions in the sky.  For
  simplicity we have assumed that the two systems have a negligible relative
  velocity when combining the two data sets.  It is clear that the addition of
  further systems will lead to a smooth coverage of the whole sky.
\label{fig:PFEarray} }
\end{figure*}

In addition, by 2013 the periastron will have advanced by nearly $\pi$ since
the pulsar discovery, which should also allow a separate measurement of
$\eta_1^{(M)}$ and $\eta_2^{(M)}$ with comparable precision. This will provide
us with measurements for $\hat{\cal Q}_1$, $\hat{\cal Q}_2$, $\hat{\cal
  Q}_1'$, and $\hat{\cal Q}_2'$, and consequently with values for the
strong-field parameters $\alpha_1^\ast$, $\alpha_2^\ast$, ${\cal A}_p$, and
${\cal A}_c$.


\section{An PFE Antenna Array}

In the future, further binary systems with a variety of orientations in space
will be discovered. In particular, with the Square-Kilometre Array (SKA) we
expect to find about 100 relativistic DNSs
(Cordes~et~al.~2004\nocite{ckl+04}; Kramer~et~al.~2004\nocite{kbc+04}). Some
of these new systems will also be suitable sources for tests of local
Lorentz-invariance of gravity, so that their observations can be combined to
derive further constraints, as also pointed out by Bailey and Kosteleck{\'y}
(2006\nocite{bk06}). Indeed, the sensitivity of additional systems to
different directions in the sky will differ from that of the Double Pulsar, so
that we can construct a dense network of antennae for studies of
preferred-frame effects. This results in a smooth and high sensitivity toward
preferred-frame effects over the whole sky.

In order to demonstrate this idea, we have simulated timing data for the
Double Pulsar system if it were located at a different position in the sky,
namely the Galactic Centre. Such a system would be sensitive to directions
which complement those of the real Double Pulsar.  This can be seen in
Figure~\ref{fig:PFEarray} where we show this situation simulated for the
timing data as expected in 2013. Comparing the left panel of this figure with
Figure 3 again demonstrates the vast improvement in the PFE limits over the
next five to seven years. The impact of the artificial system's different
position in the sky is clear from a comparison of the left and middle panel.
A combination of the two sets of limits for any given direction in the sky, as
shown in the right panel, demonstrates the concept of a PFE antenna array.

We point out that for tests of the most general theories only data from
systems with a similar combination of pulsar and companion mass can be
combined. This becomes clear when considering that the parameters ${\cal Q}_1$
and ${\cal Q}_2$ contain an explicit and implicit dependence on the neutron
star masses. The implicit dependence arises from the fact that the involved
strong-field coefficients are functions of the compactness of the bodies and
hence of the masses and the equation-of-state of the neutron stars.

If preferred-frame effects are indeed present in the timing data of a number
of binary pulsars, this array of pulsars could then be used to significantly
restrict or even determine the direction of the preferred frame (modulo $\pi$
related to a change of the sign in the measured PFE amplitudes), as a
measurement of $\tilde\chi_0$ in one such system excludes the direction to lie
within a cone with opening angle $\pi/2-\tilde\chi_0$ around the
line-of-sight to the pulsar.


\section{Summary}

We have developed a consistent methodology to measure preferred-frame effects
(PFE) related to the strong internal gravitational fields in relativistic
binary pulsars. We made only very general assumptions about the underlying
theory of gravity by using the semi-conservative generalised EIH formalism of
Will, which incorporates strong field effects in the post-Newtonian motion of
binary pulsars. We were able to show that in relativistic binary systems with
a high rate of advance of periastron, a preferred frame will cause distinctive
periodic changes in the orbital elements of the binary system.

The newly developed PFE timing model extends the DD timing model by including
these periodic changes, and thus having the amplitudes of these changes as
additional timing parameters. We described in detail how the measurement of
such PFE amplitudes can be converted into measurements of the strong field
parameters related to preferred-frame effects. 

For inclination angles $i\simeq90^\circ$, the PFE amplitudes for changes in the
longitude of periastron and the eccentricity are related by a factor which is
independent of strong field parameters and parameters related to the preferred
frame. This can be seen as a unique ``fingerprint'' of a preferred frame.

We have also shown that in the absence of preferred-frame effects, our
formalism can be used to determine limits for these amplitudes, by this
restricting the strong field parameters related to the violation of the
Lorentz invariance for nearly any direction in the sky. We demonstrated that
the Double Pulsar is the ideal test system for preferred-frame effects in
strong gravitational fields, and have presented first preliminary results,
which however are clearly less stringent then present limits from
small-eccentricity binary pulsars. On the other hand, simulations show that in
the next couple of years the precision of these tests will increase by several
orders of magnitude. The combination of several such systems in a {\em PFE
antenna array} for the detection of PFE effects can be used to obtain a full
sky coverage.


\section*{Acknowledgements}

We thank Ingrid Stairs, Dick Manchester, Maura McLaughlin, Andrew Lyne, Rob
Ferdman, Marta Burgay, Duncan Lorimer, Andrea Possenti, Nichi D'Amico, John
Sarkissian, George Hobbs, John Reynolds, Paulo Freire and Fernando Camilo for
the collaboration on the Double Pulsar. We thank Thibault Damour for many
useful and stimulating discussions, and we are in particular grateful for his
comments on the manuscript. It is also a pleasure to thank Gilles
Esposito-Far\`ese for useful comments.



\end{document}